\documentclass[aip, amsmath, amssymb, reprint]{revtex4-1}
\usepackage{graphicx}
\usepackage[english]{babel} %
\usepackage[usenames]{color}
\usepackage{subfigure}
\usepackage{dcolumn}
\usepackage{bm}
\usepackage{amsmath,amsfonts,amssymb,float}
\usepackage{natbib}
\usepackage{epsfig}
\usepackage{upgreek}
\usepackage{multirow}
\usepackage{gensymb}
\usepackage{textcomp} 
\usepackage{svg}

\makeatletter
\newcommand{\fmarki}{*}
\newcommand{\fmarkii}{\ensuremath{\dagger}}
\newcommand{\fmarkiii}{\ensuremath{\ddagger}}
\def\@fnsymbol#1{{\ifcase#1\or \fmarki\or \fmarkii\or \fmarkiii \else\@ctrerr\fi}}
\makeatother

\begin{document}
\title{Measurement of Stimulated Raman Side-Scattering Predominance in Directly Driven Experiment}
\author{K. Glize}
\thanks{These two authors contributed equally}
\author{X. Zhao}
\thanks{These two authors contributed equally}
\affiliation{Key Laboratory for Laser Plasmas (MoE) and School of Physics and Astronomy, Shanghai Jiao
Tong University, Shanghai 200240, China}
\affiliation{Collaborative Innovation Center of IFSA (CICIFSA), Shanghai Jiao Tong University,
Shanghai 200240, China}
\author{Y. H. Zhang}
\affiliation{Beijing National Laboratory for Condensed Matter Physics, Institute of Physics, Chinese
Academy of Sciences, Beijing 100190, China}
\affiliation{Collaborative Innovation Center of IFSA (CICIFSA), Shanghai Jiao Tong University,
Shanghai 200240, China}
\author{C. W. Lian}
\affiliation{Department of Plasma Physics and Fusion Engineering and CAS Key Laboratory of Geospace Environment, University of Science and Technology of China, Hefei, Anhui 230026, China}
\author{S. Tan}
\affiliation{Key Laboratory for Micro-/Nano-Optoelectronic Devices of Ministry of Education, School of Physics
and Electronics, Hunan University, Changsha, 410082, China}
\author{F. Y. Wu}
\affiliation{Key Laboratory for Laser Plasmas (MoE) and School of Physics and Astronomy, Shanghai Jiao
Tong University, Shanghai 200240, China}
\affiliation{Collaborative Innovation Center of IFSA (CICIFSA), Shanghai Jiao Tong University,
Shanghai 200240, China}
\author{C. Z. Xiao}
\affiliation{Key Laboratory for Micro-/Nano-Optoelectronic Devices of Ministry of Education, School of Physics
and Electronics, Hunan University, Changsha, 410082, China}
\author{R. Yan}
\affiliation{Department of Modern Mechanics, University of Science and Technology of China, Hefei, Anhui 230026, China}
\author{Z. Zhang}
\affiliation{Beijing National Laboratory for Condensed Matter Physics, Institute of Physics, Chinese
Academy of Sciences, Beijing 100190, China}
\affiliation{Collaborative Innovation Center of IFSA (CICIFSA), Shanghai Jiao Tong University,
Shanghai 200240, China}
\affiliation{Songshan Lake Materials Laboratory, Dongguan, Guangdong 523808, China}
\author{X. H. Yuan}
\email{Authors to whom correspondence should be addressed xiaohui.yuan@sjtu.edu.cn and jzhang1@sjtu.edu.cn}
\affiliation{Key Laboratory for Laser Plasmas (MoE) and School of Physics and Astronomy, Shanghai Jiao
Tong University, Shanghai 200240, China}
\affiliation{Collaborative Innovation Center of IFSA (CICIFSA), Shanghai Jiao Tong University,
Shanghai 200240, China}
\author{J. Zhang}
\email{Authors to whom correspondence should be addressed xiaohui.yuan@sjtu.edu.cn and jzhang1@sjtu.edu.cn}
\affiliation{Key Laboratory for Laser Plasmas (MoE) and School of Physics and Astronomy, Shanghai Jiao
Tong University, Shanghai 200240, China}
\affiliation{Collaborative Innovation Center of IFSA (CICIFSA), Shanghai Jiao Tong University,
Shanghai 200240, China}
\affiliation{Beijing National Laboratory for Condensed Matter Physics, Institute of Physics, Chinese
Academy of Sciences, Beijing 100190, China}
\begin{abstract}

Due to its particular geometry, stimulated Raman side-scattering (SRSS) drives scattered light emission in non-usually diagnosed directions, leading to scarce and complex experimental observations. Direct-irradiation campaigns at the SG-II Upgrade facility have measured the scattered light driven by SRSS over a wide range of angles. It indicated an emission at large polar angles over a broad azimuthal range, sensitive to the plasma profile and laser polarization, resulting in a loss of about 5\% of the total laser energy. Direct comparison with back-scattering measurement has evidenced SRSS as the dominant Raman scattering process. The predominance of SRSS was confirmed by two-dimensional particle-in-cell simulations, and its angular spread has been corroborated by ray-tracing simulations.
The main implication is that a complete characterization of the SRS instability and an accurate measurement of the energy losses require the collection of the scattered light in a broad range of directions. Otherwise, spatially limited measurement could lead to an underestimation of the energetic importance of stimulated Raman scattering.

\end{abstract}
\maketitle

\section{Introduction}

Despite the recent breakthrough in reaching thermonuclear fusion ignition in the laboratory \cite{Lawson_2022,Kramer_2022}, laser-plasma instabilities (LPIs) \cite{Kruer_book} remain an obstacle limiting the achievable gain in inertial confinement fusion (ICF) experiments \cite{NUCKOLLS:1972}. Stimulated Raman scattering (SRS) is a three-wave coupling resonantly driving an electron plasma wave (EPW) \cite{Comisar_1966}. This process leads to the scattering of a part of the incident laser reducing the energy coupling, and generation of a hot electron population that can preheat the fuel core. This instability is of primary concern in almost all of the ICF schemes, such as Indirect-Drive \cite{Kirkwood_2013,Montgomery2016,Hall_2017}, Direct-Drive \cite{Rosenberg_2018, Rosenberg_2020, Rosenberg_2023}, Shock Ignition \cite{Scott_2021,Cristoforetti_2019,Cristoforetti_2021,BATON_2020,Barlow_2022,Ruocco_2022} and the more recent Double-Cone Ignition \cite{Zhang_2020}. Stimulated Raman side-scattering (SRSS) is a particular SRS geometry in which the scattered light is initially emitted perpendicular to the density gradient, enabling an absolute growth (exponential growth in time at a localised spatial position) at density lower than $n_{c}/ 4$, where $n_{c}$ is the critical density. Despite extensive theoretical investigations in the late 70s \cite{Liu_1974, Mostrom_1979, Afeyan_1985}, most of the interest has been focused on stimulated Raman back-scattering (SRBS) \cite{Montgomery2016}, due to the experimental complexity to measure SRSS and the largest SRS growth rate for the backward geometry. Recently, there has been a renewed interest due to observations of SRSS on several planar direct-drive experiments \cite{Rosenberg_2018,Rosenberg_2020,Rosenberg_2023,Depierreux_2016,Depierreux_2019,Cristoforetti_2019,Ruocco_2022}, either from single beam interaction, or by multiple beams \cite{Short_2020}. Multi-beam processes happen when the laser beams are sharing a common symmetry axis enabling to drive a shared daughter wave, being either an EPW \cite{Michel_2015} or a scattered wave \cite{Depierreux_2016, Depierreux_2019}. These experimental observations have led to the development of a more complete analytical description of the SRSS, accounting for the convective nature (finite spatial amplification while propagating through the resonant region) of the instability near the turning point, in order to explain the SRSS growth in region below the absolute threshold \cite{Michel_2019}. It highlighted that ICF experiments are prone to being SRSS unstable as the instability can extend to lower densities in the convective regime due to the large dimensions of the interaction, namely long density scale-length, large laser focal spot and high temperature. Therefore a complete understanding of this detrimental process is imperative as leading to additional laser energy coupling loss and hot electron generation. However, experimental observations have been restricted to a limited number of directions since ICF large laser facilities are usually not designed to measure SRS in directions other than back-scattering. Thus, comprehensive measurements of this mechanism and related losses are still not available. In order to improve the overall characterisation of SRSS, crucial new diagnostics are currently being implemented in order to provide further observations at additional angles of observations \cite{Rosenberg_RSI_2021,Trauchessec_2022}.

In this manuscript, we present a highly-resolved 2D angular measurement of SRS evidencing the importance of side-scattering in Direct-Drive ICF experiments, even at shorter density scale-length usually considered Raman stable conditions \cite{Seka_2009,Seka_2014}. During these experimental campaigns, a new diagnostic has been designed and implemented to provide a measurement of the SRS emission spectrum over several directions: Angular-Resolved Scattered-light Diagnostic Station (ARSDS) \cite{Zhao_RSI_2021}. Direct comparison with the back-scattered light collected in the aperture of one of the driver beam evidenced SRSS as the dominant SRS process. 2D PIC simulations confirmed that in this regime, the interaction was below the threshold for the typical SRBS to grow and only SRSS was responsible for the SRS emission. Due to its broad angular emission, SRSS was identified to be responsible for the scattering of at least $5\pm2\%$ of the total laser energy and requires to be measured over broad range of directions. Ray-tracing simulations have been performed, confirming SRSS as the origin of the measured broadband emission with such large angular spread. 

\section{Experimental setup}
\begin{figure}
\centering 
\includegraphics[width=0.480\textwidth]{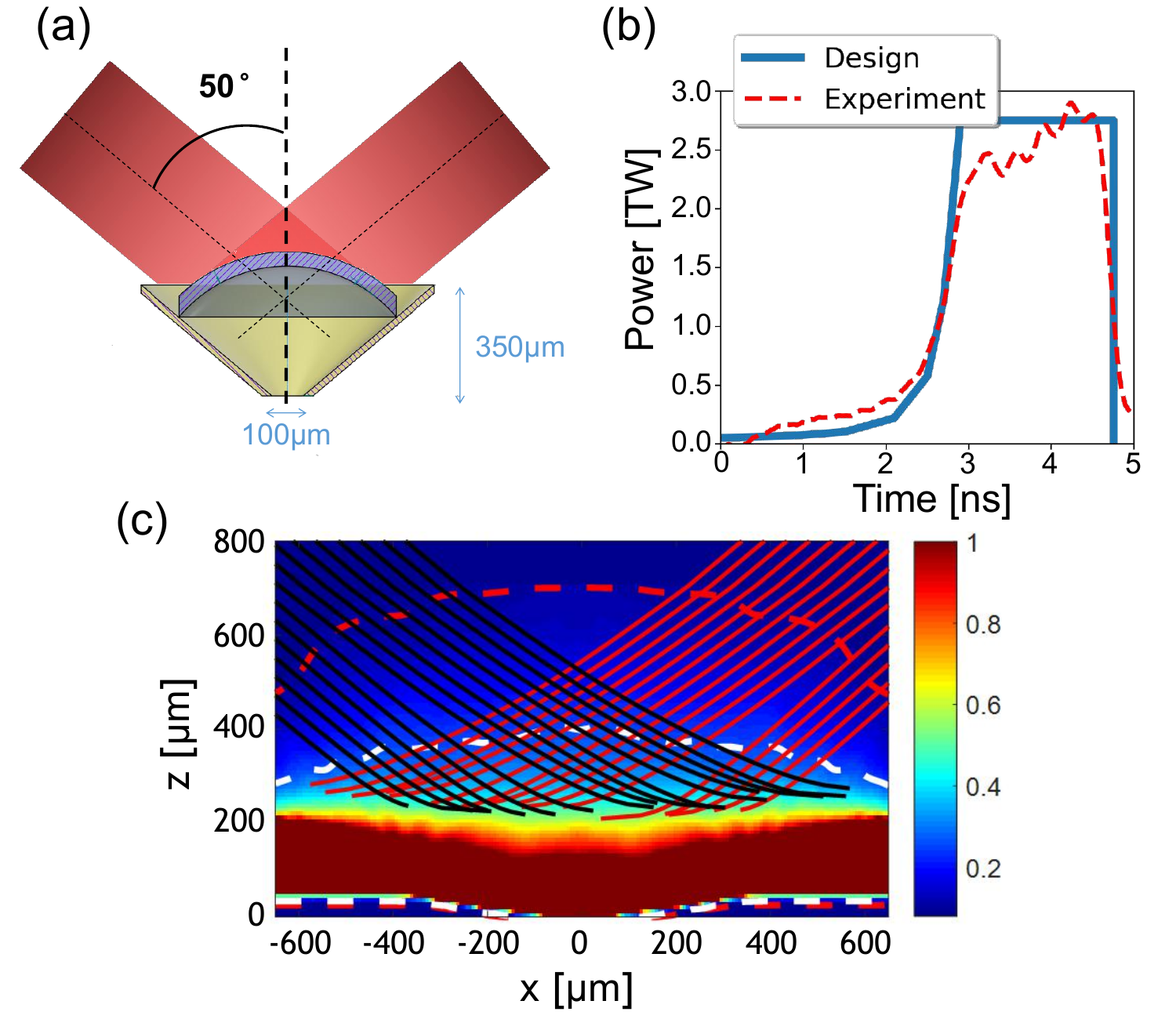}
\caption{\label{Setup} (a) Schematic of the irradiation geometry on the spherical cap target. (b) Pulse shape of the overlapped power delivered on target, both by design (blue) and experimentally (dashed red). (c) Density map of the planar target at $t = 4.25~$ns, with $n_{e} = 0.01~n_{c}$ and $0.25~n_{c}$ isocontour in red and white dashed lines, respectively. Only two laser beams, in black and red rays, are shown for clarity.}
\end{figure}

The experiment was performed at the SG-II Upgrade laser facility \cite{SGIIU}, which provides eight nanosecond pulses organised in two cones of four beams, propagating from the upper and lower hemispheres, respectively. 
Our experiment focused on the upper hemisphere interaction, using the setup displayed on Fig. \ref{Setup}(a). A cone of four beams, \#1, \#3, \#5 and \#7, were used to irradiate the target at a polar angle of $\theta=50 \degree$, with a uniform distribution in the azimuthal direction. For clarity, we define here that beam \#7 azimuthal position is the origin, $\varphi=0 \degree$, of the azimuth axis and that the upper pole is the origin, $\theta=0 \degree$, of the polar axis. Each $310 \times 310~$mm square beam delivers 1.5 kJ at 351 nm focused by a $f/7.1$ wedged lens on target. The beams were CPP smoothed to obtain a 525$~\upmu$m ($1/e^{2}$) focal spot. The pulse shape consisted of a 2.9~ns exponential ramp, followed by a 1.8~ns plateau, as depicted on Fig. \ref{Setup}(b). This corresponds to a peak overlapped intensity on target of $\approx 1.2 \times 10^{15}~$W.cm$^{-2}$ in vacuum. The beams are primarily $p$-polarised plus a rotation angle of $7 \degree$ for beams \#3, \#5 and $23 \degree$ for beams \#1, \#7.

Planar and segmented spherical CH targets have been used, which are commonly considered in Direct-Drive related experiments. The planar targets consisted of a 50$~\upmu$m-thick C16H16 layer, backed with a 200$~\upmu$m-thick SiO2 layer. The spherical targets consisted of a  45$~\upmu$m-thick, 450$~\upmu$m inner radius C16H16 spherical cap, contained within a 20$~\upmu$m-thick, open-ended Au cone in order to maintain a spherical compression \cite{Zhang_2020}. This particular targets were designed and used in the context of the Double-Cone Ignition approach \cite{Zhang_2020}. It was verified that the ablation depth for our conditions was $\leq 30~\upmu$m \cite{Liu_2022}. At peak power, plasma parameters were estimated using the hydro-radiative code MULTI2D \cite{RAMIS2009}, with an Arbitrary Lagrangian-Eulerian module. Spherical and Cartesian grids have been used for the spherical cap and planar target simulations, respectively. It predicts a density scale-length of $L_{n_{c}/ 4} \approx 250 ~\upmu$m (resp. $\approx 175 ~\upmu$m) at $n_{c}/ 4$ up to $L_{n_{c}/ 10} \approx 350 ~\upmu$m (resp. $\approx 250 ~\upmu$m) in the coronal plasma for planar target (resp. spherical target), with a uniform electron temperature from $T_{e} \approx 2.0$ to $2.5~$keV. The electron temperature near $n_{c} / 4$ was inferred to be $\leq 2.2~$keV from the red-shifted spectral feature related to Two-Plasmon Decay (TPD) instability \cite{Seka_1985}, as shown by the inset in Fig. \ref{SRSS_spectrum}(a). 

Three diagnostics have been used to characterise laser-plasma instabilities. First, a full-aperture back-scattering station (FABS) was installed on \#7 in order to collect the light scattered in the backward direction $[\theta, \varphi]=[50\degree, 0\degree]$. The back-scattered light was collected on the transmission of the last dichroic mirror, and focused by a  $360 \times 360~$mm lens on the diagnostic table. The signal was divided into three arms, to measure the spectrum, energy and temporal shape, using a Coherent energy-meter, an Ocean-Optics spectrometer and a Si photodiode, respectively. Both energy and temporal measurement were filtered in spectrum using a OG550 Schott filter. A Lambertian diffuser was placed before both the spectral and temporal measurement in order to suppress any achromatic aberration impact on the signal. Second, the ARSDS diagnostic  was used to collect the scattered light at the coordinates depicted by the red circles in Fig. \ref{60F} and measure the temporally integrated SRS spectra resolved in angle for both azimuthal and polar axes. It consists of three arrays of fibers placed 50~cm away from the target, in the proximity of \#7. All fibers, with low attenuation in the relevant spectral domain, are then coupled to an iHR500 imaging spectrometer, in order to provide a SRS spectrum with high-angular resolution. A $533 \pm 8.5~$nm notch filter was used to eliminate any residual 527 nm light and a $400~$nm long-pass filter was used to minimize the effect of the second-order diffraction of the main laser around 702~nm. Last, 40 additional fibers \cite{Zhang_RSI_2021} were placed on the inner wall of the spherical target chamber, at a distance of 1.2~m from the target, at the coordinates depicted by the white diamonds in Fig. \ref{60F}. This diagnostic have been used to collect SRS light in further directions, over a larger angular domain with lower angular resolution. Its setup was similar to ARSDS, with all fibres being coupled to an Isoplane160 imaging spectrometer. All three diagnostics have been absolutely calibrated in energy using a white LED source of known emission placed at the target chamber centre.

\section{Experimental results}

\begin{figure}
\centering 
\includegraphics[width=0.5\textwidth]{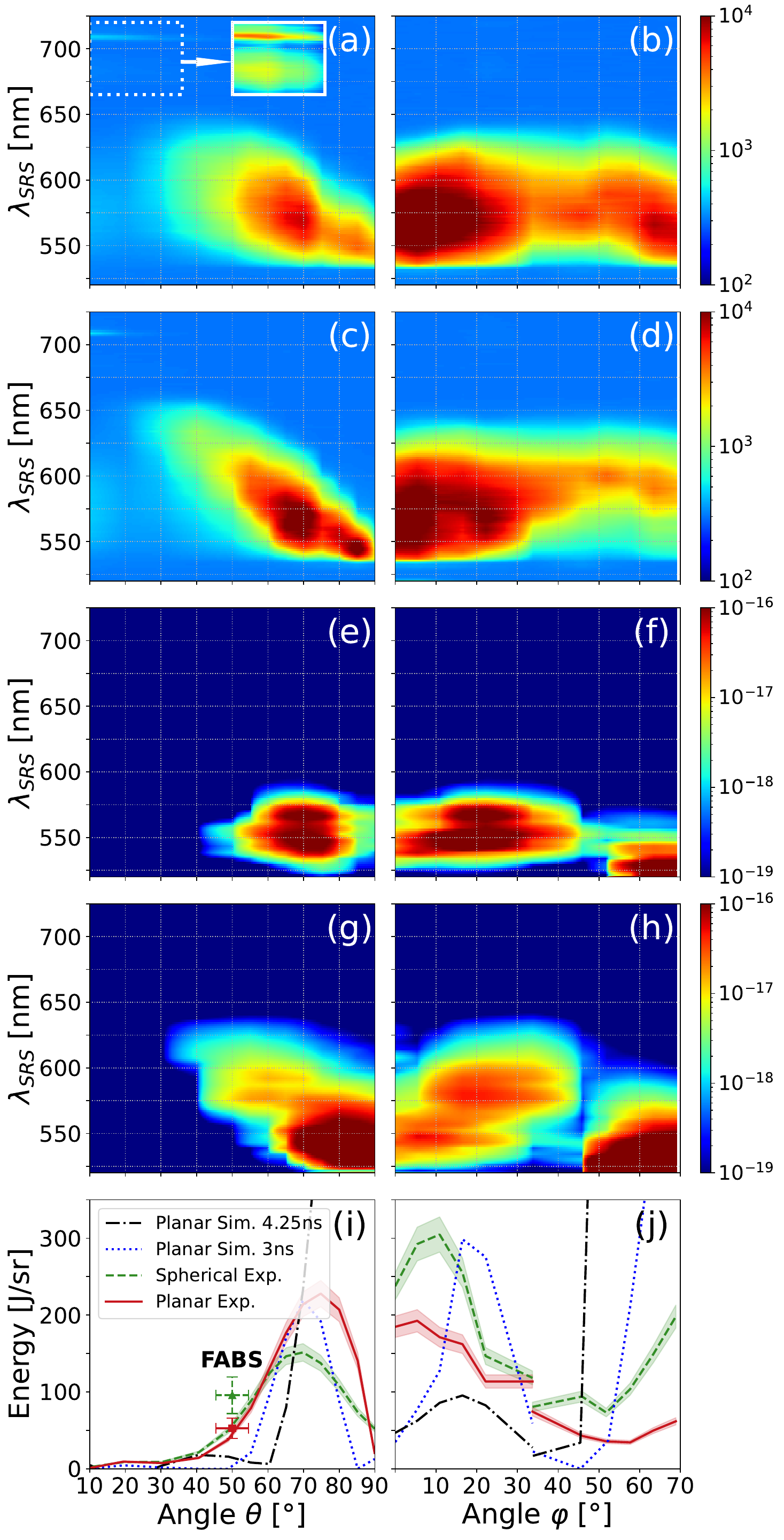}
\caption{\label{SRSS_spectrum} Typical SRS spectrum measured by the ARSDS diagnostic, for spherical and planar targets, respectively, against (a, c) the polar angles collected at an azimuthal angle of $33.7 \degree$; (b, d) the azimuthal angles collected at polar angles displayed by the red circles in Fig. \ref{60F}. The inset in (a) presents the contrast enhanced TPD features. (e),(f) and (g),(h) are the spectrum against the polar and azimuthal angles at $t=3~$ns and $t=4.25~$ns, respectively, from the ray-tracing simulation considering a planar target. (i) and (j), green dashed and red plain lines are the scattered energy against the exit angle, integrated over the respective spectrum and normalized to the collection solid angle, for the the spherical and planar experiments, respectively. Back-scattered energy measured by FABS is displayed as a green triangle and a red square on (i). The discontinuity in (j) is due to the different polar coordinates, as shown in Fig. \ref{60F}. The blue dotted lines and black dash-dot lines present the scattered light signal in arbitrary units from the ray-tracing simulations considering a planar target, at $t=3~$ns and $t=4.25~$ns, respectively.}
\end{figure}

A typical result is presented on Fig. \ref{SRSS_spectrum}, displaying the scattered light spectrum angularly resolved in the (a, c) polar and (b, d) azimuthal directions for spherical and planar targets, respectively. In both cases, the overall signal is dominated by light with a spectrum ranging from 530~nm up to 640~nm detected within a wide range of directions from $30 \degree$ up to $90 \degree$ polar angles, over the whole azimuth, corresponding to a density range of $ \approx [0.1-0.2]n_{c}$. The green dashed and red plain curves on Figs. \ref{SRSS_spectrum}(i) and \ref{SRSS_spectrum}(j) present, respectively, the signal from Figs. \ref{SRSS_spectrum}(a)-\ref{SRSS_spectrum}(d) integrated over the spectrum, calibrated in energy and normalised to the solid angle. It shows that most of the scattered light is detected at large polar angles, peaked around $\theta = 70 \degree$ for the spherical target, and $\theta = 75 \degree$ in the planar case. Such signal corresponds to an SRS emission at shorter wavelength and larger angle than previously reported, where usually the collection of SRS light is limited to angles $\leq 50 \degree$. Furthermore, Fig. \ref{SRSS_spectrum}(j) shows that the amount of scattered energy is sensitive to the azimuthal position. For the spherical case, the maximum emission is not contained in the azimuthal plane of the laser beam but is offset by an angle $\Delta \varphi \approx 10\degree$. 
Considering that the broadband signal is detected at polar angles larger than the incident driver beams, over a large azimuthal section, it evidences that side-scattering is responsible for such emission. The SRSS scattered light is emitted orthogonal to the density gradient and experiences refraction on its way out of the plasma \cite{Afeyan_1985,Michel_2019}. This results in a correlation between the scattered light wavelength and the polar exit angle, as observed on Figs. \ref{SRSS_spectrum}(a) and \ref{SRSS_spectrum}(c). Moreover, SRSS scattered light is also predominantly emitted perpendicular to the laser polarization plane, as reported in early experiment \cite{Drake_1984} and confirmed by theoretical \cite{Kruer_book,Afeyan_1985,Menyuk_1985} and numerical \cite{Xiao_2016, Xiao_2018,Gu_2021} studies. 
From these spectra, it appears that the scattering is driven by single beam SRSS, as multi-beam process would drive scattered light either in the bisector plane at $\varphi \approx$ 45$\degree$ for a shared scattered electromagnetic wave \cite{Depierreux_2016, Depierreux_2019}, or constrained to density $n_{e} \leq 0.12n_{c}$ for a shared EPW due to the large angle between two neighbouring beams \cite{Michel_2015}. This was further confirmed by experimentally measuring a two-orders-of-magnitude decrease of the signal in the polar direction when \#7 is switched off. Azimuthal measurement was not available for this test. Thus, single-beam intensity is considered in the following discussion. However, it is important to mention that the interaction geometry is not favourable to characterize multi-beam scattering. Indeed, the laser beams are mostly overlapping at density higher than $n_{c}/4$, as depicted in Fig. \ref{Setup}(c), limiting the growth of shared waves, and thus being a suitable platform for single beam instability observation. Besides, difference in plasma profiles between the ideal spherical compression from the simulation \cite{Yang_2022} and the actual experiment have been observed from preliminary angular filter refractometry measurement \cite{Haberberger_2014} (not presented here). It appeared that the plasma is accumulating close to the Au cone, leading to a flattening of the coronal plasma, in agreement with the similarities in SRSS emission for both spherical and planar targets. Thus, only the planar case will be considered in the simulations and discussions section, due to the sensitivity of the SRSS emission to the plasma profile.

\begin{figure}
\centering 
\includegraphics[width=0.40\textwidth]{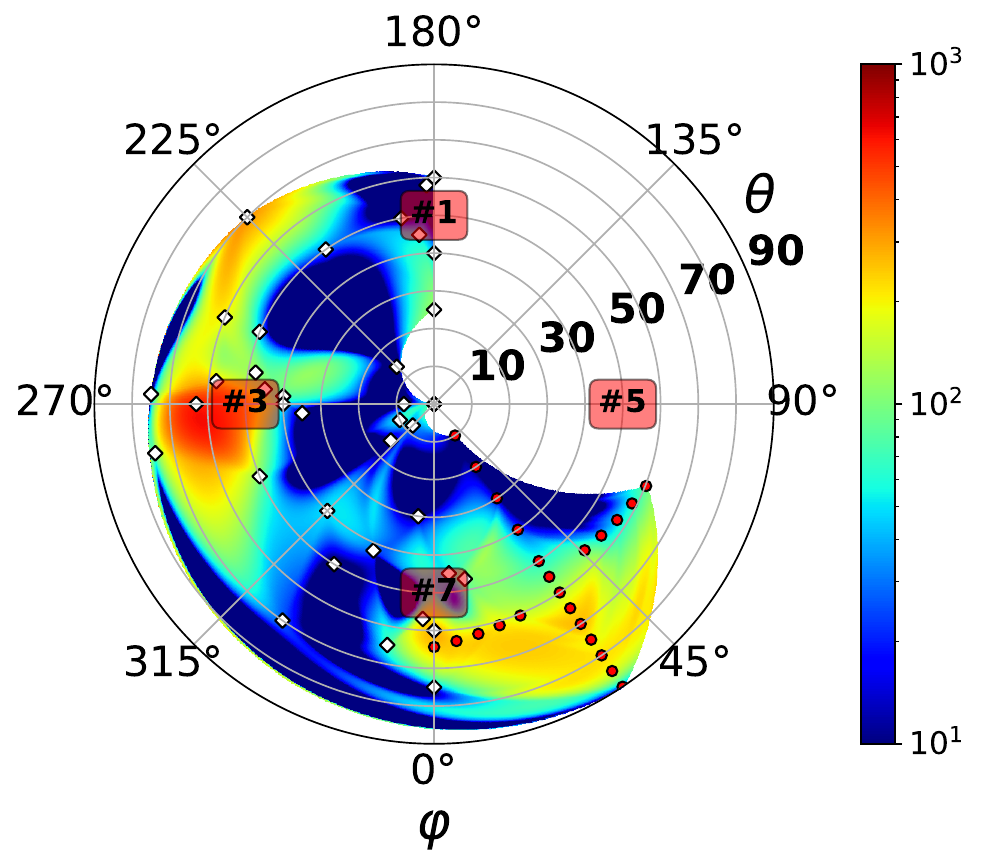}
\caption{\label{60F} Interpolation of the scattered energy distribution in J/sr, integrated in time and spectrum, from the signal measured by FABS on beam \#7, ARSDS (red circles) and additional fibers (white diamonds), for the same experiment as displayed in Figs. \ref{SRSS_spectrum}(c) and \ref{SRSS_spectrum}(d) using a planar target.}
\end{figure}

FABS measurements typically show a similar spectrum and amount of scattered energy than ARSDS (up to $\approx1.5$ higher), for the same polar angle as shown on the Fig. \ref{SRSS_spectrum}(i), despite the difference in azimuth, consistent with Fig. \ref{SRSS_spectrum}(j). Such measurement evidences that there is no stronger emission localised in the backward direction which could be attributed to SRBS. This is expected as the laser intensity is one order of magnitude lower than the predicted SRBS threshold of $\approx~3\times10^{15}~$W.cm$^{-2}$ for our conditions \cite{Estabrook1983, Kruer_book, Montgomery2016}. Such conditions also prevents SRBS to grow from the high intensity speckles \cite{Rousseaux2016, Glize2017}, which have been inferred to reach up to $\approx 2.5 \times 10^{15}~$W.cm$^{-2}$ . This results in a negligible reflectivity measured by FABS, $\leq 0.15\pm 0.05$\%, as being only a small fraction of the total scattered energy. In order to account for the large scattering angular spread, using the 40 additional fibers along with ARSDS and FABS enabled to extend the measurement over a $\frac{4}{3} \pi$ solid angle. Despite the limited resolution and inherent uncertainties due to the finite number of directions probed, the scattered energy distribution can be interpolated as displayed on Fig. \ref{60F} for the planar target experiment. First, it appears that there is no stronger signal at polar angle $\theta \leq 50 \degree$ in the azimuthal plane $\varphi \approx 0 \degree$ of \#7. This is confirming the absence of SRBS signal, which could have experienced refraction and not be completely collected by FABS. Second, a reflectivity of $5\pm2\%$ of the total laser energy was estimated. However, this is only a low boundary  estimation as the absorption of the SRS light in the plasma before being collected is not considered. Indeed, an accurate estimation of the total energy losses of the driver laser beams requires to estimate the spectrally dependent absorption experienced by the light collected by each detectors. Furthermore, as the detected light is dependent on the actual plasma profile, and that the plasma is expanding over time, the dynamic absorption for each detectors needs to be assessed for the whole interaction duration. Lastly, due to the absence of signal detected at wavelength higher than $640~$nm, it is impossible to accurately determine SRSS losses in this spectral range. Nonetheless, this evidences that SRSS is energetically significant, even in our short density scale-length conditions where SRS was usually assumed stable \cite{Seka_2009,Seka_2014}, and necessitates to be measured over the whole angular domain with sufficient sensitivity. This implies that SRSS may lead to additional losses not usually accounted for and could be one candidate to explain some ``missing'' energy reported in recent experiments \cite{Hall_2017,Turnbull_2022}. Indeed, it appears that SRSS losses can be easily overlooked as only scattering a small amount of energy locally, which would normally not be detected nor considered.

\section{Simulations and Discussions}

\begin{figure}
\centering 
\includegraphics[width=0.5\textwidth]{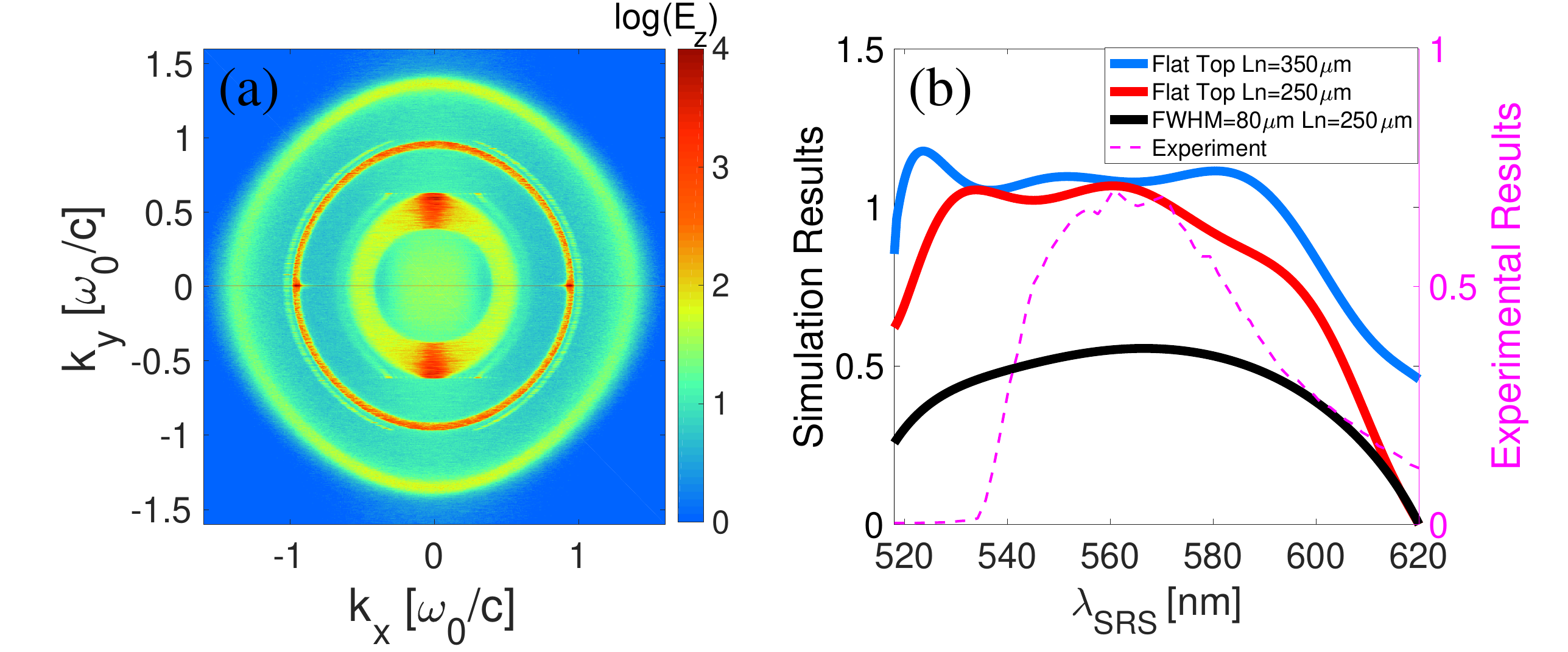}
\caption{\label{PIC_result} (a) 2D spectrum of the scattered light electric field $E_{z}$ averaged over 2~ps for $L_{n} = 350 ~\upmu$m using a plane wave, (b) and associated wavelength from PIC simulations in the blue curve. Red and black lines present the scattered light wavelength for $L_{n} = 250 ~\upmu$m with a plane wave and a Gaussian wave receptively. The purple dashed lines present the signal from Fig. \ref{SRSS_spectrum}(c) integrated over the angles.}
\end{figure}

First, 2D plane-wave PIC simulations have been performed using the code EPOCH \cite{Arber_2015} to confirm the predominance of SRSS over SRBS and that SRSS is responsible for most of the emission measured experimentally. The full simulation box is $250 ~\upmu$m in length (x-axis) and $100 ~\upmu$m in width (y-axis), with a $10 ~\upmu$m vacuum at the left boundary. The longitudinal boundary conditions are open for the fields and thermal for the particles, and the transverse boundary condition is periodic. The spatial resolution is $0.05~\upmu$m in both directions and the temporal resolution is $20~$fs. Each cell initially contains 60 particles. An $s$-polarized plane wave, with an intensity of  $I_0=4.5\times10^{14}~$W.cm$^{-2}$, is normally incident into a linear-density-profile plasma slab ranging from $0.1n_c$ to $0.2n_c$, with a density scale-length of $350 ~\upmu$m, an electron temperature $T_e=2~$keV and fixed ions. Figure \ref{PIC_result}(a) shows a time-averaged 2D spectrum of the electromagnetic wave $E_{z}$ over $2~$ps. The brightest signals are the scattered light near $90 \degree$, perpendicular to the density gradient. As expected, there is no measurable SRBS. The blue curve on Fig. \ref{PIC_result}(b) is a conversion into wavelength of the scattered light spectrum from Fig. \ref{PIC_result}(a). It confirms SRSS growth over the whole range of density simulated, consistent with the experiment displayed by the the purple dashed line ($i.e$ Fig. \ref{SRSS_spectrum}(c) integrated over the angles), despite an extension to lower wavelength for the simulation. Additional simulations with $L_{n} = 250 ~\upmu$m consistent with the spherical case, and with a finite Gaussian beam of $80 ~\upmu$m FWHM, have also been performed. These show identical conclusions and similar spectra as depicted by the red and black curves, respectively. The broader spectra in the PIC simulations as compared to the experimental results could be attributed to larger inherent numerical seeds in PIC simulations.

Second, analytical description can be considered to investigate SRSS growth regime in our experimental conditions. In ICF experiments, SRSS can experience both absolute or convective growth, depending on the interaction conditions. In our conditions, considering oblique incidence and damping, the absolute threshold \cite{Afeyan_1985} is overcome for $\lambda_{SRS} \geq 550~$nm, due to low $T_{e}$. The lower density part can be interpreted considering the convective regime \cite{Michel_2019}, having a gain $G \geq 7$ for the spectral range observed experimentally. The absence of signal above 640~nm is likely due to the significant re-absorption at the associated high density, as previously reported \cite{Depierreux_2016, Michel_2019}. At low density, the finite size of the beam is limiting the convective growth of SRSS \cite{Michel_2019}, which requires a large transverse amplification length. This corresponds to a cutoff around 530~nm for a 525$~\upmu$m diameter focal spot, consistent with our experimental observations.

Last, 3D ray-tracing simulations, using the code PHANTAM \cite{Ji_2022}, have been performed to explore the origin of the observed angular spread of the scattered light. Indeed, such investigation is quite complex, due to the 2D plasma expansion, the convective nature of the instability, the inherent sensitivity of SRSS to the plasma profile and laser polarization; and thus requires dedicated numerical tools \cite{Hironaka_2023}. This 3D code, based on the method published by Kaiser \cite{Kaiser_2000}, has been developed to simulate the propagation of the SRSS light, accounting for the convective gain \cite{Michel_2019}, the collisional absorption and the Landau damping. The simulation box is $0.4\times0.4\times0.3~$cm with a $200\times200\times300$ mesh, containing a plasma obtained from the hydrodynamic simulations. The simulation domain includes all four beams, with their respective polarization, and an intensity of $I_0=4.5\times10^{14}~$W.cm$^{-2}$. They are propagating into a plasma obtained from the MULTI2D hydrodynamic simulations at the time step $t=3~$ns, corresponding to the early stage of the laser power plateau were SRSS emission is starting to reach its maximum level, as observed when coupling one of the ARSDS fiber to a spectrometer resolved in time. SRSS rays are initialized perpendicular to both the local density gradient, with a 9$\degree$ random spreading angle accounting for near-tangential emission, and the local polarization of the incident ray. Each ray initial intensity is determined by the local convective gain \cite{Michel_2019}. Upon exiting the plasma, SRSS rays are collected at the same angles as the ARSDS detectors. Figs. \ref{SRSS_spectrum}(e) and \ref{SRSS_spectrum}(f) display the angularly resolved spectrum from the simulation, which appear to qualitatively reproduce the experimental spectrum on Figs. \ref{SRSS_spectrum}(c) and \ref{SRSS_spectrum}(d), respectively. One can notice a slight discrepancy for wavelength $\geq 590~\upmu$m, where the experimental signal appears stronger than the simulation. This could be due to the absolute nature of the instability at higher wavelength which is not accounted for in the simulation. It was observed that modifying the electron temperature from $T_{e} = 2~$keV to $2.5~$keV, was changing the lower wavelength emission boundary from $520~$nm to $530~$nm. This serves as an evidence that the Landau cutoff also limits the emission at lower wavelength. By integrating the signal over the spectrum, the blue dotted curves in Figs. \ref{SRSS_spectrum}(i) and \ref{SRSS_spectrum}(j) are confirming the overall scattering angular profile, with peaked signals at large polar angles, maximized near the beam azimuth. In the simulation, the emission is peaked for an azimuth $\varphi \approx 20\degree $, which was observed experimentally for the spherical target, and is most likely due to difference in the plasma profile between the experiment and the hydro-simulation. 
Ray-tracing simulation in a plasma profile at the time step $t=4.25~$ns, towards the end of the laser pulse, shows an SRSS emission at higher wavelength as displayed on Figs. \ref{SRSS_spectrum}(g) and \ref{SRSS_spectrum}(h), in closer agreement with the experimental data.  However, the angular scattering profile appears to be quite different, as shown by the black dash-dot curves on Figs. \ref{SRSS_spectrum}(i) and \ref{SRSS_spectrum}(j), mostly due to a very strong emission at lower wavelength. Considering that the experimental measurement are time integrated, it evidences the necessity to integrate over multiple time steps in order to improve the matching of the numerical results. However, due to the computing time cost and the storage space required for a single simulation, this is currently beyond our capacity.
By alternatively and independently switching on and off each beam, it was observed that the two beams \#5 and \#7 are predominantly responsible for the detected signal. For the azimuthal detectors, the signal close to \#7 at $\varphi \leq 25 \degree$ is actually originating from \#5. The remaining part of the signal at larger azimuthal angle is predominantly driven by \#7. For the polar detectors, the peak signal at large polar angle is due to beam \#7 only. This is in agreement with the drastic reduction of the signal measured when \#7 is switched off. This seems to indicate that near-tangential scattering is responsible for the measured signal. However, recent simulations of OMEGA EP experiments evidenced the contribution of SRSS at non-tangential angles \cite{Hironaka_2023}. Similar method was applied in our case but were leading to results radically different than the experimental observations, with an emission peaked at 600~nm in most directions. This highlights the complexity of SRSS, and the necessity of further highly resolved experimental measurements to benchmark and improve predictive capabilities. Further investigations, beyond the scope of this paper, are necessary to improve numerical agreement with the experiment and comprehensively understand SRSS mechanism, and will be the subject of future work.

\section{Conclusions}
In conclusion, the importance of stimulated Raman side-scattering in a regime usually considered SRS stable, as below the back-scattering threshold, has been experimentally observed and confirmed by 2D PIC simulations. Due to the sensitivity to both plasma profile and laser polarization, SRSS scattered light is emitted over a broad range of directions, as observed experimentally and confirmed by 3D ray-tracing simulations. Such broad angular scattering results in a small amount of energy being scattered locally, while being energetically significant when measured over the whole angular domain. Thus, our results are highlighting the necessity to use diagnostic methods with large angular detection and high sensitivity to accurately measure SRS activity. Indeed, relying on limited directions to diagnose SRS, such as back-scattering, could lead to a strong underestimation of the actual energy losses, by a factor of at least 35 in our conditions. 
However, these result have been obtained at reduced plasma conditions compared to ignition-scale Direct-Drive experiments. Thus, further investigations are required to assess the impact of SRSS under these conditions, such as: (i) the competition between Raman side- and back-scattering when SRBS threshold is overcome; (ii) the effect of laser smoothing techniques such as smoothing by spectral dispersion (SSD) \cite{Skupsky1989} which could reportedly mitigate SRSS \cite{Kang2021}; (iii) SRSS behavior at higher intensities relevant to the Shock Ignition approach \cite{Betti_2007}. 

\begin{acknowledgments}
We thank the SG-II Upgrade laser facility operating group and target fabrication team for their assistance. This work was supported by the Strategic Priority Research Program of Chinese Academy of Sciences (grants No. XDA25010100, No. XDA25030200, No. XDA25050700 and No. XDA25050400).
\end{acknowledgments}

\section*{AUTHOR DECLARATIONS}
\subsection*{Conflict of Interest}
The authors have no conflicts to disclose.

\section*{Data Availability Statement}
The data that support the findings of this study are available from the corresponding author upon reasonable request.

\bibliography{./Biblio.bib}

\end{document}